\documentclass[referee]{raa}            

\usepackage{graphicx,times}             
\usepackage[a4paper,margin=1in]{geometry}
\usepackage{array}
\usepackage{natbib}
\usepackage{orcidlink}
\usepackage{amssymb,amsmath}
\usepackage{CJK}
\usepackage[T1]{fontenc}
\usepackage[utf8]{inputenc}
\usepackage[english]{babel}
\usepackage[autopunct=true]{csquotes}
\bibpunct{(}{)}{;}{a}{}{,}

\usepackage{hyperref}

\begin{document}
\begin{CJK*}{UTF8}{gbsn}

  \title{The Pantheon Sample Analysis of Cosmological Constraints under New Models
}

   \volnopage{Vol.0 (20xx) No.0, 000--000}      
   \setcounter{page}{1}          

   \author{Peifeng Peng (彭培风)\,\orcidlink{0000-0002-8486-621X} 
      \inst{1}
   \and Nigel Metcalfe\,\orcidlink{0000-0001-9034-4402}
      \inst{2}
   }

   \institute{Department of Physics, Washington University, 1 Brookings Dr, St. Louis, MO 63130, USA; {\it p.peifeng@wustl.edu}\\
        \and
             Department of Physics, Durham University, South Road, Durham DH1 3LE, UK\\
\vs\no
   {\small Received 20xx month day; accepted 20xx month day}}

\abstract{In this paper, the cosmological parameters are determined by applying six cosmological models to fit the magnitude-redshift relation of the Pantheon Sample consisting of 1048 Type Ia supernovae (SNe Ia) in the range of $0.01 < z < 2.26$. Apart from the well-known flat $\Lambda$CDM model as well as other models that have been broadly studied, this paper includes two new models, the $ow$CDM model and the $ow_{0}w_{a}$CDM model, to fully evaluate the correlations between the cosmological parameters by performing the MCMC algorithm and to explore the geometry and mass content of the Universe. Combining the measurements of the baryon acoustic oscillation (BAO) and the cosmic microwave background (CMB) with the SNe Ia constraints, the matter density parameter $\Omega_\mathrm{M} = 0.328^{+0.018}_{-0.026}$, the curvature of space parameter $\Omega_{k} = 0.0045^{+0.0666}_{-0.0741}$, and the dark energy equation of state parameter $w = -1.120^{+0.143}_{-0.185}$ are measured for the $ow$CDM model. When it comes to the $ow_{0}w_{a}$CDM model, if the parameter $w$ is allowed to evolve with the redshift as $w = w_{0} + w_{a}\left(1-a\right)$, the cosmological parameters are found to be $\Omega_\mathrm{M} = 0.344^{+0.018}_{-0.027}$, $\Omega_{k} = 0.0027^{+0.0665}_{-0.0716}$, $w_{0} = -0.739^{+0.336}_{-0.378}$, and $w_{a} = -0.812^{+0.750}_{-0.678}$. The $ow$CDM model and the $ow_{0}w_{a}$CDM model are able to fit the Pantheon Sample consistently well with $\chi_{\nu}^{2} = 0.994$ and $\chi_{\nu}^{2} = 1.008$, but the parameters $w_{0}$ and $w_{a}$ are not well constrained in both models. Meanwhile, the flat $\Lambda$CDM model is found to fit poorly for $z > 0.5$ high-redshift SNe Ia data with $\chi_{\nu}^{2} = 0.792$ compared to the $w_{0}w_{a}$CDM model with $\chi_{\nu}^{2} = 0.971$ and the $ow_{0}w_{a}$CDM model with $\chi_{\nu}^{2} = 0.824$.
}

   \authorrunning{P. Peng \& N. Metcalfe}            
   \titlerunning{Determining Cosmological Parameters by New Models}  

   \maketitle

%
%
\section{Introduction}           
\label{section1}

Since Edwin Hubble discovered the first observational evidence of the expanding Universe and derived his renowned Hubble's Law \citep{1929PNAS...15..168H}, our understanding of the Universe has been completely reshaped. Hubble's Law formed the basis of modern observational cosmology, which supported the Big Bang model for the origin of our Universe and led to the research of large-scale structures, distant galaxy clusters, and supernovae \citep{doi:10.1073/pnas.1424299112}. Among all these candidates, Type Ia supernovae (SNe Ia) arouse the interest of cosmologists the most because they share consistent peak luminosity, meaning that they can be used as standard candles to accurately measure their distances to us and hence determine the cosmological constraints \citep{1938ApJ....88..285B, 1998AJ....116.1009R}.

The Supernova Cosmology Project was started about 35 years ago, with the first significant work observed and collected 60 SNe Ia data at that time to study their magnitude-redshift relation \citep{1999ApJ...517..565P}. From
the deviation of linearity in the SNe Ia Hubble diagram at $z \sim 0.5$, it was surprisingly found that the expansion rate of our Universe was currently accelerating due to the existence of dark energy which accounted for ${\sim}70\%$ of the total energy in the Universe \citep{1998AJ....116.1009R, 1999ApJ...517..565P, 2001ApJ...560...49R}. Under a flat $\Lambda$CDM model, the cosmological parameters were determined as $\Omega_\mathrm{M} = 0.28$ and $\Omega_\mathrm{\Lambda} = 0.72$ \citep{1999ApJ...517..565P}.

In recent years, as more SNe Ia were measured by the Pan-STARRS1 (PS1) Survey \citep{2016arXiv161205560C}, the fitting ability of the flat $\Lambda$CDM model needs to be reviewed. Nowadays, modern cosmological probes have suggested that the flat $\Lambda$CDM model cannot fully describe our Universe \citep{2016A&A...594A..13P}. Under the assumption of other cosmological models, this paper aims at reanalyzing and constraining the cosmological parameters to better fit the magnitude-redshift relation of the SNe Ia data.

The supernovae dataset used in this paper is based on the Pantheon Sample \citep{2018ApJ...857...51J, 2018ApJ...859..101S}, which is a full set of 1048 SNe Ia data in the range of $0.01 < z < 2.26$ combined from different surveys, including PS1 \citep{2016arXiv161205560C}, the Supernova Legacy Survey \citep[SNLS;][]{2011ApJS..192....1C}, the Sloan Digital Sky Survey \citep[SDSS;][]{2009ApJS..185...32K}, and the Hubble Space Telescope Survey \citep[HST;][]{2007ApJ...659...98R}. The SNe Ia data of the Pantheon Sample have all been calibrated and standardized \citep{2018ApJ...857...51J, 2018ApJ...859..101S} with the systematic uncertainties evaluated, such as the photometric calibration, the Milky Way extinction, the mass estimates for the host galaxies of SNe Ia, etc. The full list of 1048 corrected SNe Ia table can be viewed at \href{https://archive.stsci.edu/doi/resolve/resolve.html?doi=10.17909/T95Q4X}{DOI: 10.17909/T95Q4X}.

In this paper, the derivation and physical interpretation of the cosmological models and the MCMC algorithm are explained in Section~\ref{section2}. Then, the flat $\Lambda$CDM model is applied in Section~\ref{section3} to fit the magnitude-redshift relation of the Pantheon Sample, which will prove to be a relatively poor fit for high-redshift SNe Ia data when $z > 0.5$. Hence, other cosmological models, including two new models $ow$CDM and $ow_{0}w_{a}$CDM, will all be applied in Section~\ref{section3} to better fit the Pantheon Sample by performing the MCMC algorithm so that the correlations between the cosmological parameters can be fully analyzed with their uncertainties estimated. The MCMC corner plots of all models are illustrated in Section~\ref{section3}, Appendix~\ref{appendixa}, and Appendix~\ref{appendixb}. Finally, there will be a short conclusion given in Section~\ref{section4} of this paper.


\section{Methods}
\label{section2}

Firstly, the Friedmann equation is expressed below, which governs the evolution of a homogeneous isotropic Universe under general relativity by relating the expansion rate to the energy density \citep{1999GReGr..31.1991F}:
\begin{equation}
    H^{2} = \left(\frac{\dot{a}}{a}\right)^{2} = \frac{8\pi G}{3c^{2}}\left(\rho_\mathrm{mass}c^{2} + \rho_\mathrm{DE}c^{2}\right) - \frac{kc^{2}}{a^{2}}\,,\label{equation1}
\end{equation}
where $H$ is the Hubble parameter, $a$ is the scale factor, $G$ is the gravitational constant, $c$ is the speed of light, $\rho_\mathrm{mass}c^{2}$ is the energy density of matter, $\rho_\mathrm{DE}c^{2}$ is the energy density of dark energy, and $k$ is the curvature of space. Due to the conservation law of matter in the Universe, the density of matter follows as $\rho_\mathrm{mass} = \rho_{0}a^{-3}$, where $\rho_{0}$ is the present-day density of matter. Meanwhile, the density of dark energy can be derived by $\rho_\mathrm{DE} = \rho_\mathrm{\Lambda, 0}a^{-3\left(1+w\right)}$, where $\rho_{\mathrm{\Lambda, 0}}$ is the present-day density of dark energy, $w = P_\mathrm{DE}/\rho_\mathrm{DE}c^{2}$ is the dark energy equation of state parameter defined from the fluid equation by assuming our Universe as an expanding fluid. Thus, let $\Omega_\mathrm{M,0} = 8\pi G\rho_{0}/3H_{0}^{2}$ and $\Omega_\mathrm{\Lambda,0} = 8\pi G\rho_\mathrm{\Lambda,0}/3H_{0}^{2}$, the original Friedmann equation can be rewritten as:
\begin{equation}
    \frac{H^{2}}{H_{0}^{2}} = \frac{\Omega_\mathrm{M,0}}{a^{3}} + \frac{\Omega_\mathrm{\Lambda,0}}{a^{3\left(1+w\right)}} + \frac{\Omega_{k,0}}{a^{2}}\,,\label{equation2}
\end{equation}
where $\Omega_\mathrm{M,0}$ is the present-day matter density parameter, $\Omega_\mathrm{\Lambda,0}$ is the present-day dark energy density parameter, $\Omega_{k,0} = -kc^{2}/H_{0}^{2}$ is the present-day curvature of space parameter, and $H_{0}$ is the Hubble constant. Note that the Hubble constant cannot be directly estimated from the PS1 Survey alone because it requires more distance indicators, such as the Cepheid variables or the CMB and BAO experiments based on the Wilkinson Microwave Anisotropy Probe \citep[WMAP;][]{2013ApJS..208...20B}, to constrain $H_{0}$ by combining multiple surveys \citep{2016A&A...594A..13P, 2022ApJ...938..110B}. Therefore, the prior value of the Hubble constant is taken to be $H_{0} = 70\,\mathrm{km}\,\mathrm{s}^{-1}\,\mathrm{Mpc}^{-1}$ in this paper after considering the \textquote{Hubble tension} between the local expansion rate ($H_{0}$) measurements and the early Universe predictions \citep{2016ApJ...826...56R, 2022ApJ...938..110B}.

Under the flat $\Lambda${CDM} model, the density of dark energy $\rho_\mathrm{DE}$ is invariant to the scale factor $a$, implying that $\Omega_\mathrm{\Lambda, 0}$ is a constant value, which is also known as the cosmological constant. Furthermore, since the Universe is currently in the $\Lambda$-dominated era under the flat $\Lambda$CDM model, two conditions $\Omega_{k, 0} = 0$ and $w = -1$ must be satisfied, and the present-day radiation density parameter $\Omega_\mathrm{R, 0}$ can be safely ignored. As a result, the simple relation $\Omega_\mathrm{M, 0} + \Omega_\mathrm{\Lambda, 0} = 1$ for the flat $\Lambda$CDM model can then be obtained from equation~\ref{equation2} by assuming $H = H_{0}$ and $a = 1$ for the present-day values.

\begin{figure}
  \centering
  \includegraphics[width=1.0\linewidth]{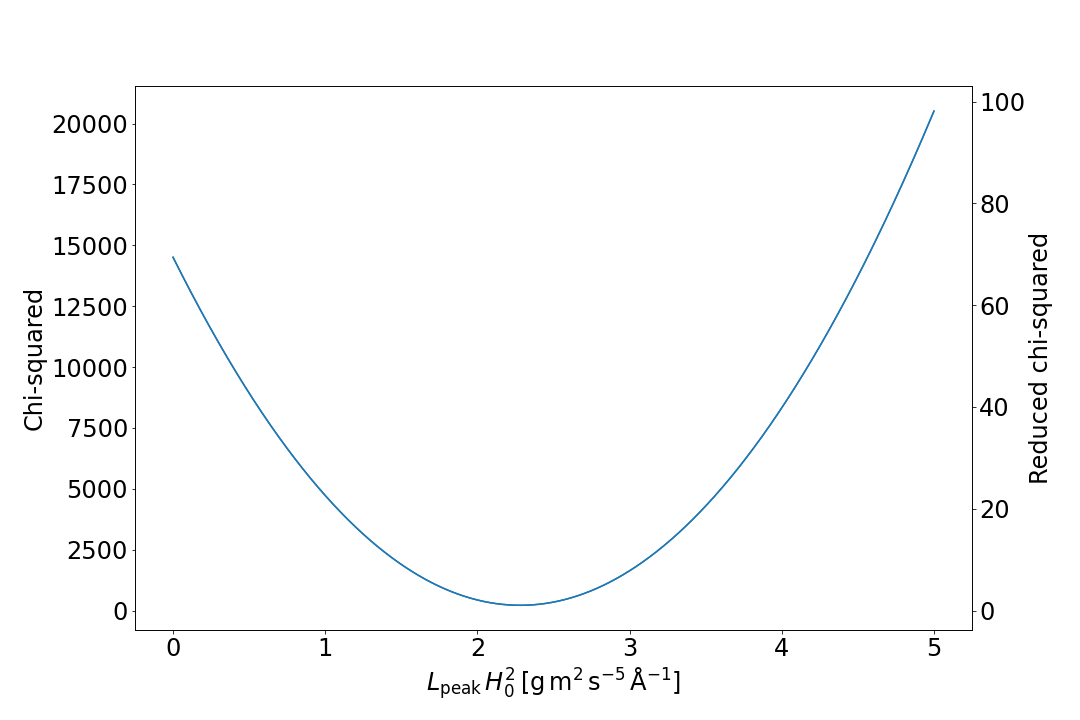}
  \caption{The variation of chi-squared ($
  \chi^{2}$) and reduced chi-squared ($\chi_{\nu}^{2}$) with $L_\mathrm{peak}H_{0}^{2}$ after performing the chi-squared minimization analysis. The local minimum corresponds to the best-fit $L_\mathrm{peak}H_{0}^{2}$.}
  \label{figure1}
\end{figure}

To calculate the effective magnitude $m$ of each Pantheon Sample SNe Ia data, one needs to find the best-fit peak luminosity $L_\mathrm{peak}$, which is related to the detected flux $f$ by:
\begin{equation}
    f = \frac{L_\mathrm{peak}}{4\pi S_{k}\left(\eta\right)^{2}\left(1+z\right)^{2}}\,,\label{equation3}
\end{equation}
where $z$ is the redshift of each Pantheon Sample SNe Ia data with heliocentric and peculiar velocity corrections performed by \citet{2022PASA...39...46C}, $\eta$ is the comoving distance, and $S_{k}\left(\eta\right)$ is the Friedmann-Robertson-Walker (FRW) metric. The detected flux $f$ can be further converted to the effective magnitude $m$ using:
\begin{equation}
    m = m_{B} - 2.5\log_{10}f\,,\label{equation4}
\end{equation}
with $m_{B} = -20.48$ the rest-frame B band peak magnitude \citep{Marchesini_2007}. For the low-redshift SNe Ia data ($z < 0.1$), the approximation $S_{k}\left(\eta\right) = \eta = cz/H_{0}$ is valid by assuming a flat $\Lambda$CDM model. Then, the best-fit peak luminosity $L_\mathrm{peak}$ can be determined via a chi-squared minimization analysis by looping $L_\mathrm{peak}H_{0}^{2}$ from 0 to 5 over 1,000 steps in the unit of $\mathrm{g}\,\mathrm{m}^{2}\,\mathrm{s}^{-5}\, \mathring{\mathrm{A}}^{-1}$. The outcome is shown in Figure~\ref{figure1}, where the best-fit $L_\mathrm{peak}H_{0}^{2}$ at the local minimum is equivalent to $L_\mathrm{peak} = \left(4.43^{+0.04}_{-0.03}\right) \times 10^{39}\,\mathrm{erg}\,\mathrm{s}^{-1}\,\mathring{\mathrm{A}}^{-1}$. The uncertainties of the best-fit $L_\mathrm{peak}$ are estimated within $\pm1\sigma$ (68.3\%) confidence interval, which yields $\Delta \chi^{2}
= \pm1$ for one degree of freedom \citep{Hughes_2010}.

After obtaining the best-fit $L_\mathrm{peak}$, the comoving distance $\eta$ can be accurately computed as\footnote{The integral is solved using the \textit{scipy.integrate} package in Python}:
\begin{equation}
    \eta = \frac{c}{H_{0}\sqrt{\Omega_{k}}}S_{k}\left[\sqrt{\Omega_{k}}\int_{0}^{z}\frac{dz'}{E(z')}\right]\,,\label{equation5}
\end{equation}
and $E\left(z'\right)$ is given by \citep{2018ApJ...857...51J, 2018ApJ...859..101S}:
\begin{equation}
    E\left(z'\right) = \left[\Omega_\mathrm{M}\left(1+z'\right)^{3} + \Omega_\mathrm{\Lambda}\left(1+z'\right)^{3\left(1+w\right)} + \Omega_{k}\left(1+z'\right)^{2}\right]^{\frac{1}{2}}\,,\label{equation6}
\end{equation}
Note that for simplicity, the parameter $\Omega_\mathrm{M}$, $\Omega_\mathrm{\Lambda}$, and $\Omega_{k}$ in equation~\ref{equation6} are equivalent to the present-day value $\Omega_\mathrm{M, 0}$, $\Omega_\mathrm{\Lambda, 0}$, and $\Omega_{k, 0}$ mentioned above. The function $S_{k}\left(x\right)$ in equation~\ref{equation5} follows the conditions $S_{k}\left(x\right) = \sin{x}$ when $\Omega_{k} < 0$, $S_{k}\left(x\right) = x$ when $\Omega_{k} = 0$, and $S_{k}\left(x\right) = \sinh{x}$ when $\Omega_{k} > 0$. Similarly, the FRW metric $S_{k}\left(\eta\right)$ in equation~\ref{equation3} is defined by the expressions below:
\begin{equation}
S_k\left(\eta\right) =
\begin{cases}
    \frac{\sin{\sqrt{k}\eta}}{\sqrt{k}}, & k > 0 \\
    \eta, & k = 0 \\
    \frac{\sinh{\sqrt{-k}\eta}}{\sqrt{-k}}, & k < 0\,\label{equation7}
\end{cases}
\end{equation}

To solve the integral in equation~\ref{equation5}, one could still use the chi-squared minimization analysis to find the best-fit cosmological parameters for the flat $\Lambda$CDM model. However, when it comes to other more complicated models (such as the $ow$CDM model and the $ow_{0}w_{a}$CDM model), it is extremely time-consuming to minimize chi-squared in higher dimensional parameter space. Thus, a computational algorithm called the affined invariant Markov Chain Monte Carlo (MCMC) is performed here to efficiently calculate multiple best-fit cosmological parameters and estimate their uncertainties at the same time by fully evaluating the correlations in the parameter space. Built from the \textit{emcee} \citep{2013PASP..125..306F} and \textit{corner} \citep{Foreman-Mackey2016} package in Python, the MCMC algorithm can automatically draw samples from the posterior probability distribution of parameters where the advance of each step in the Markov Chain is only dependent on the location of its previous step \citep{2013PASP..125..306F}. Constrained by the prior function and the likelihood function, the sampled parameters describe the highest likelihood model. The likelihood function $P\left(X|\Theta,\alpha\right)$ is determined by a chi-squared test, which only keeps the parameters in the MCMC algorithm that pass this test:
\begin{equation}
    P\left(X|\Theta,\alpha\right) = \frac{1}{2}\sum_{i}\left(\frac{Y_{i}-Y\left(X_{i}\right)}{\sigma_{i}}\right)^{2}\,,\label{equation8}
\end{equation}
where $X$ and $Y$ are the independent variable and the set of observations (dependent variable), $\sigma$ is the set of observation errors, $\alpha$ is the set of nuisance parameters to realize the marginalization process, and $\Theta$ is the set of model parameters. The highest likelihood model is generated by the advances of the parameter vector which follow the likelihood function $P\left(X|\Theta,\alpha\right)$ within the prior function $P\left(\Theta,\alpha\right)$. Specifically speaking, the range of each parameter in the prior function $P\left(\Theta,\alpha\right)$ is set to be broader enough to fully take into account the correlations in the parameter space. The posterior probability distribution of parameters $P\left(\Theta,\alpha|X\right)$ is therefore derived to be proportional to the product of $P\left(X|\Theta,\alpha\right)$ and $P\left(\Theta,\alpha\right)$ in accordance with Bayes' Theorem \citep{2013PASP..125..306F}:
\begin{equation}
   P\left(\Theta, \alpha|X\right) \sim P\left(\Theta, \alpha\right)P\left(X|\Theta, \alpha\right)\,,\label{equation9}
\end{equation}
After performing the MCMC algorithm over 2,000 steps for a 1-hour sampling process, the best-fit cosmological parameters under different cosmological models can be obtained to fit the magnitude-redshift relation of the Pantheon Sample. Another outcome of the MCMC algorithm is the corner plots, which can also be illustrated to evaluate the correlations between the best-fit parameters and estimate their uncertainties. At last, a chi-squared hypothesis test containing the reduced chi-squared ($\chi_{\nu}^{2}$) and the P-value is performed to evaluate the fitting ability of different cosmological models when fitting the Pantheon Sample.

\section{Results and Discussion}
\label{section3}

\begin{figure*}
  \begin{minipage}{0.5\textwidth}
     \centering
     \includegraphics[width=1.0\linewidth]{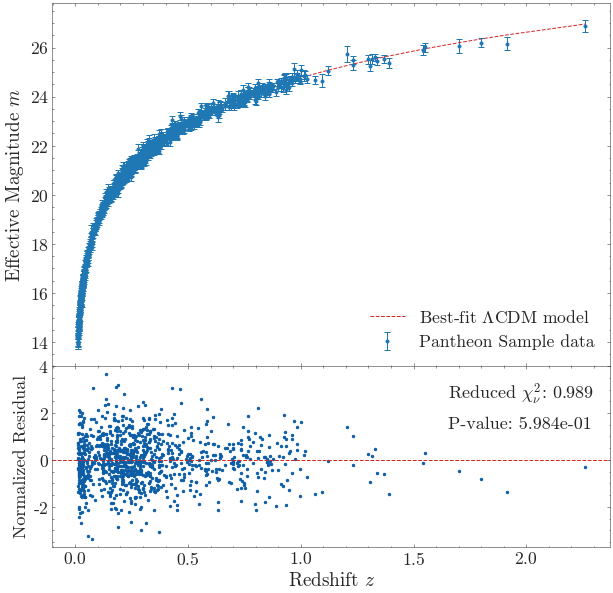}
   \end{minipage}\hfill
   \begin{minipage}{0.5\textwidth}
     \centering
     \includegraphics[width=1.0\linewidth]{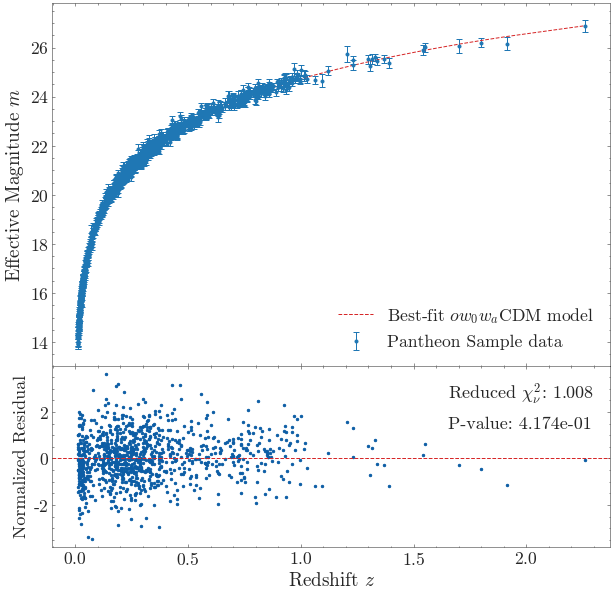}
   \end{minipage}
  \caption{The left panel is the magnitude-redshift relation of the Pantheon Sample fitted by the flat $\Lambda$CDM model when $\Omega_\mathrm{\Lambda} = 0.703 \pm 0.016$ and $\Omega_\mathrm{M} = 0.281 \pm 0.011$, while the right panel is the same diagram fitted by the $ow_{0}w_{a}$CDM model when $\Omega_\mathrm{\Lambda} = 0.644^{+0.185}_{-0.145}$ and $\Omega_\mathrm{M} = 0.341^{+0.033}_{-0.051}$. The best-fit values and the uncertainties of $\Omega_\mathrm{\Lambda}$ and $\Omega_\mathrm{M}$ are constrained by running the MCMC algorithm for 2,000 steps. The normalized residual plot is included underneath to help visualize the fitting ability of these two models.}
  \label{figure2}
\end{figure*}

\begin{figure*}
  \centering
  \includegraphics[width=1.0\linewidth]{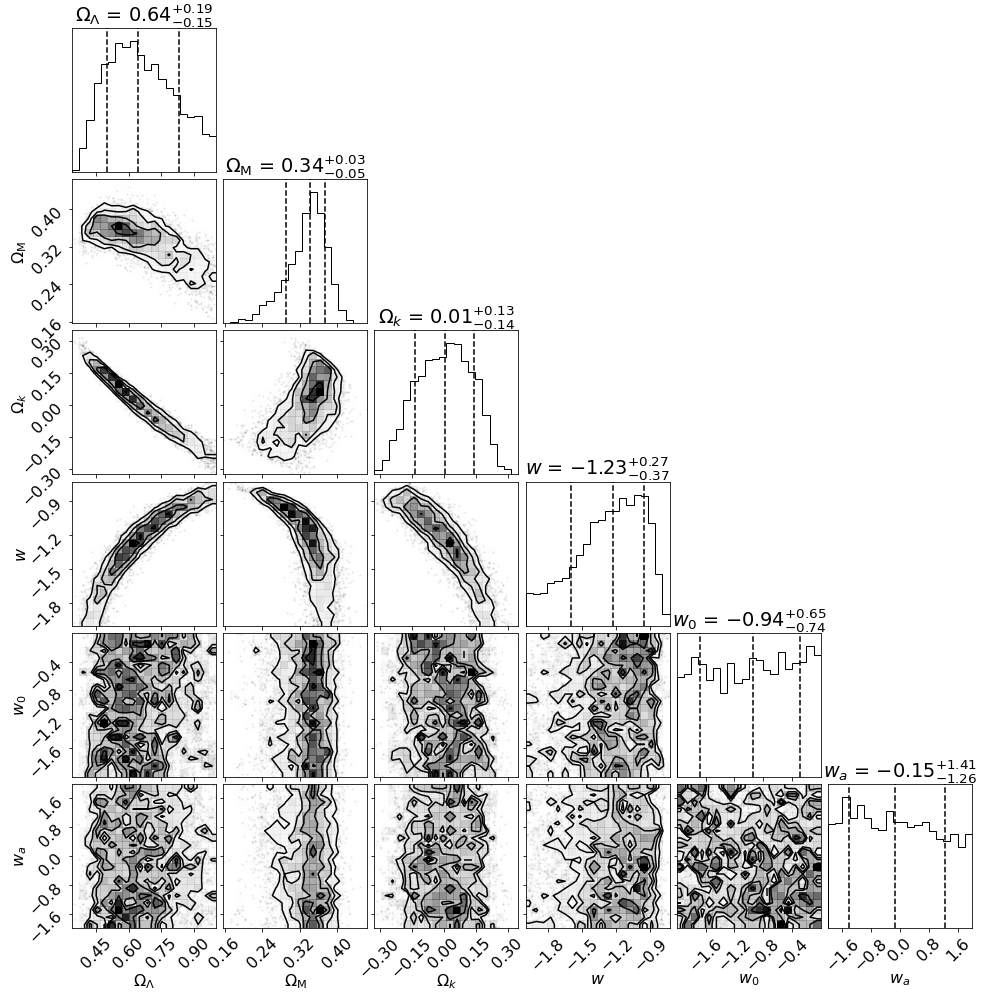}
  \caption{The MCMC corner plot for the $ow_{0}w_{a}$CDM model. The best-fit cosmological parameters ($\Omega_\mathrm{\Lambda}, \Omega_\mathrm{M}, \Omega_{k}, w, w_{0}, w_{a}$) and their uncertainties are determined by running the MCMC algorithm for 2,000 steps. It is obviously seen that two parameters $w_{0}$ and $w_{a}$ are poorly constrained in this corner plot, indicating that the MCMC algorithm must be improved in the future.}
  \label{figure3}
\end{figure*}

\renewcommand{\arraystretch}{1.0}
\begin{table*}
  \begin{center}
    \begin{tabular}{>{\centering\arraybackslash}p{1.8cm} >{\centering\arraybackslash}p{2.5cm} >{\centering\arraybackslash}p{2.0cm} >{\centering\arraybackslash}p{2.0cm} >{\centering\arraybackslash}p{2.0cm} >{\centering\arraybackslash}p{2.0cm}}
    \hline\hline
    Model & Measurements & $\Omega_\mathrm{M}$ & $\Omega_{k}$ & $w_{0}$ & $w_{a}$ \\
    \hline
        $\Lambda$CDM & PS1 & $0.281 \pm 0.011$ & / & / & / \\
        $\Lambda$CDM & PS1+CMB+BAO & $0.296 \pm 0.006$ & / & / & / \\
        $o$CDM & PS1 & $0.308 \pm 0.019$ & $-0.119 \pm 0.068$ & / & / \\
        $o$CDM & PS1+CMB+BAO & $0.309 \pm 0.010$ & $-0.059 \pm 0.034$ & / & / \\
        $w$CDM & PS1 & $0.350^{+0.033}_{-0.039}$ & / & $-1.235^{+0.138}_{-0.143}$ & / \\
        $w$CDM & PS1+CMB+BAO & $0.329^{+0.017}_{-0.020}$ & / & $-1.122^{+0.072}_{-0.074}$ & / \\
        $ow$CDM & PS1 & $0.343^{+0.035}_{-0.051}$ & $0.0079^{+0.1332}_{-0.1481}$ & $-1.246^{+0.281}_{-0.366}$ & / \\
        $ow$CDM & PS1+CMB+BAO & $0.328^{+0.018}_{-0.026}$ & $0.0045^{+0.0666}_{-0.0741}$ & $-1.120^{+0.143}_{-0.185}$ & / \\
        $w_{0}w_{a}$CDM & PS1 & $0.351^{+0.033}_{-0.037}$ & / & $-0.947^{+0.649}_{-0.715}$ & $0.010^{+1.368}_{-1.349}$ \\
        $w_{0}w_{a}$CDM & PS1+CMB+BAO & $0.349^{+0.019}_{-0.021}$ & / & $-0.751^{+0.336}_{-0.368}$ & $-0.645^{+0.728}_{-0.719}$ \\
        $ow_{0}w_{a}$CDM & PS1 & $0.341^{+0.033}_{-0.051}$ & $0.0073^{+0.1329}_{-0.1433}$ & $-0.942^{+0.652}_{-0.737}$ & $-0.153^{+1.412}_{-1.259}$ \\
        $ow_{0}w_{a}$CDM & PS1+CMB+BAO & $0.344^{+0.018}_{-0.027}$ & $0.0027^{+0.0665}_{-0.0716}$ & $-0.739^{+0.336}_{-0.378}$ & $-0.812^{+0.750}_{-0.678}$ \\
        \hline
    \end{tabular}
    \caption{The best-fit cosmological parameters ($\Omega_\mathrm{M}, \Omega_{k}, w_{0}, w_{a}$) and their uncertainties for six models obtained by combining different measurements after running the MCMC algorithm for 2,000 steps. The \textquote{/} sign represents that the parameter is not defined under that model.}
    \label{table1}
  \end{center}
\end{table*}

\begin{figure}
  \centering
  \includegraphics[width=1.0\linewidth]{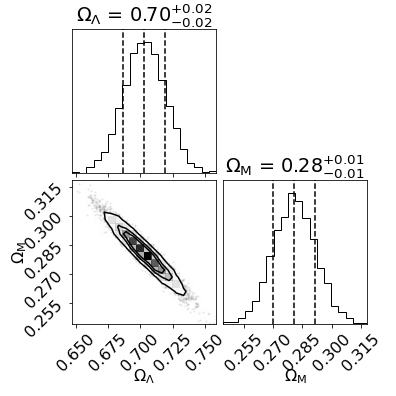}
  \caption{The MCMC corner plot for flat $\Lambda$CDM model. The best-fit cosmological parameters ($\Omega_\mathrm{\Lambda}, \Omega_\mathrm{M}$) and their uncertainties are determined by running the MCMC algorithm for 2,000 steps. Both parameters ($\Omega_\mathrm{\Lambda}, \Omega_\mathrm{M}$) are well constrained in this corner plot.}
  \label{figure4}
\end{figure}

In this paper, six cosmological models are explored: the flat $\Lambda$CDM model ($\Omega_{k} = 0$, $w = -1$), the $o$CDM model ($\Omega_{k}$ varies, $w = -1$), the flat $w$CDM model ($\Omega_{k} = 0$, $w$ varies), the owCDM model ($\Omega_{k}$ varies, $w$ varies but not evolves with $a$, $w_{a} = 0$), the flat $w_{0}w_{a}$CDM model ($\Omega_{k} = 0$, $w$ varies and evolves with a under the relation $w = w_{0} + w_{a}\left(1 - a\right)$, $w_{0}$ and $w_{a}$ vary), and the $ow_{0}w_{a}$CDM model ($\Omega_{k}$ varies, $w_{0}$ and $w_{a}$ vary). All models are performed by the MCMC algorithm to determine the best-fit parameters and estimate their uncertainties, while only the flat $\Lambda$CDM model, the $ow_{0}w_{a}$CDM model, and the $ow$CDM model are plotted in Figure~\ref{figure2} and in the left panel of Figure~\ref{figure_a1} to fit the magnitude-redshift relation of the Pantheon Sample with the normalized residual plot included underneath. Their corresponding MCMC corner plots are then illustrated in Figure~\ref{figure3}, Figure~\ref{figure4}, and in the right panel of Figure~\ref{figure_a1} to show how the parameters are correlated with each other through the contour plots.

In combination with the CMB \citep{2016A&A...594A..13P} and BAO measurements, the results of all best-fit cosmological parameters under different models are presented in Table~\ref{table1}. The BAO measurements are taken
from the SDSS Data Release 7 (DR7) main galaxy sample \citep{2015MNRAS.449..835R}, SDSS-III Baryon Oscillation Spectroscopic Survey (BOSS) DR 10, 11, and 12 \citep{2013AJ....145...10D, 2014MNRAS.441...24A, 2017MNRAS.470.2617A}. To validate the reliability of the best-fit cosmological parameters under the $o$CDM model, the flat $w$CDM model, and the flat $w_{0}w_{a}$CDM model, their corresponding MCMC corner plots are also illustrated in Figure~\ref{figure_b1} and Figure~\ref{figure_b2} in Appendix~\ref{appendixb}.

\renewcommand{\arraystretch}{1.0}
\begin{table*}
  \begin{center}
    \begin{tabular}{>{\centering\arraybackslash}p{2.0cm} >{\centering\arraybackslash}p{2.0cm} >{\centering\arraybackslash}p{2.5cm} >{\centering\arraybackslash}p{2.0cm} >{\centering\arraybackslash}p{2.5cm}}
    \hline\hline
    Model & $\chi_{\nu}^{2}$ & $\chi_{\nu}^{2}$ ($z > 0.5$) & P-value & P-value ($z > 0.5$) \\
    \hline
        $\Lambda$CDM & 0.989 & 0.792 & 0.598 & 0.989 \\
        $o$CDM & 0.988 & 0.788 & 0.603 & 0.991 \\
        $w$CDM & 0.985 & 0.787 & 0.628 & 0.991 \\
        $ow$CDM & 0.994 & 0.796 & 0.553 & 0.988 \\
        $w_{0}w_{a}$CDM & 1.056 & 0.971 & 0.103 & 0.608 \\
        $ow_{0}w_{a}$CDM & 1.008 & 0.824 & 0.417 & 0.973 \\
        \hline
    \end{tabular}
    \caption{The reduced chi-squared ($\chi_{\nu}^{2}$) and the P-value of six models.}
    \label{table2}
  \end{center}
\end{table*}

Based on the results of the chi-squared hypothesis test demonstrated in Table~\ref{table2}, all cosmological models (except the $w_{0}w_{a}$CDM model) applied in this paper are good fits for the Pantheon Sample since the $\chi_{\nu}^{2} \rightarrow 1$ and the P-value $\rightarrow 0.5$ \citep{Hughes_2010}. However, from the normalized residual plot in the left panel of Figure~\ref{figure2}, high-redshift SNe Ia data when $z > 0.5$ are slightly deflected from the central horizontal line, revealing that the flat $\Lambda$CDM model becomes a poor fit for $z > 0.5$ SNe Ia data and hence cannot fully describe the geometry and mass content of our Universe in the past, which is supported by the faintness in the SNe Ia Hubble diagram suggesting that our Universe starts to accelerate at $z \sim 0.5$ \citep{1998AJ....116.1009R, 1999ApJ...517..565P, 2001ApJ...560...49R}. Furthermore, This finding is validated by calculating the $\chi_{\nu}^{2}$ and the P-value of all cosmological models for $z > 0.5$ SNe Ia data in Table~\ref{table2}, which demonstrates that both the $w_{0}w_{a}$CDM model and the $ow_{0}w_{a}$CDM model have a closer $\chi_{\nu}^{2}$ and P-value to their ideal values (1 and 0.5) than the flat $\Lambda$CDM model.

Regarding the results listed in Table~\ref{table1}, all the best-fit cosmological parameters are consistent with the previous literature results \citep{2018ApJ...857...51J, 2018ApJ...859..101S} because they all agree with a Universe which contains ${\sim}30\%$ baryonic and non-baryonic matter and ${\sim}70\%$ dark energy \citep{doi:10.1073/pnas.1424299112}, although the parameter $\Omega_\mathrm{M}$ is slightly overestimated for the models apart from the flat $\Lambda$CDM model and the $o$CDM model. Besides, the uncertainties of two parameters $w_{0}$ and $w_{a}$ for the flat $w_{0}w_{a}$CDM model and the $ow_{0}w_{a}$CDM model are extremely large compared to other parameters, causing them to be poorly constrained by the MCMC algorithm and hence less reliable. These poor constraints are also reflected in the MCMC corner plot in Figure~\ref{figure3}, where two parameters $w_{0}$ and $w_{a}$ fail to construct valid contour plots. To further constrain the cosmological parameters and reduce their uncertainties, one could either run the MCMC algorithm for more steps or perform other MCMC algorithms that converge faster, such as the one built from the \textit{pymcmcstat} package in Python \citep{Miles2019}.

Even though the flat $\Lambda$CDM model has been proven to be a relatively poor fit when fitting $z>0.5$ SNe Ia data, the general best-fit parameters presented in Table~\ref{table1} indicate that the flat $\Lambda$CDM model is still convincing because $\Omega_{k} \approx 0$, inferring that our Universe is almost spatially flat at present. Meanwhile, since the best-fit parameter $w \neq -1$, it also suggests that the parameter $w$ is likely to evolve with time as $w = w_{0} + w_{a}\left(1 - a\right)$, which sheds light on the importance of investigating the best-fit results of two parameters $w_{0}$ and $w_{a}$ under new cosmological models \citep{2016A&A...594A..13P}.

\section{Conclusions}
\label{section4}

In this paper, the magnitude-redshift relation of the Pantheon Sample consisting of 1048 SNe Ia data is fitted by six cosmological models, including two new models the $ow$CDM model and the $ow_{0}w_{a}$CDM model, to remeasure the cosmological parameters. After performing the MCMC algorithm and illustrating the MCMC corner plots, the correlations in the parameter space can be evaluated through the contour plots to determine the best-fit cosmological parameters and estimate their uncertainties. When determining the best-fit parameters, the CMB and BAO measurements are combined with the SNe Ia data to improve the accuracy of the cosmological constraints, giving the results of the matter density parameter $\Omega_\mathrm{M} = 0.328^{+0.018}_{-0.026}$, the curvature of space parameter $\Omega_{k} = 0.0045^{+0.0666}_{-0.0741}$, and the dark energy equation of state parameter $w = -1.120^{+0.143}_{-0.185}$ for the $ow$CDM model. Similarly, the results for the $ow_{0}w_{a}$CDM model yield as $\Omega_\mathrm{M} = 0.344^{+0.018}_{-0.027}$, $\Omega_{k} = 0.0027^{+0.0665}_{-0.0716}$, $w_{0} = -0.739^{+0.336}_{-0.378}$, and $w_{a} = -0.812^{+0.750}_{-0.678}$.

The best-fit parameters of all models applied in this paper are compatible with the literature results, except that the flat $\Lambda$CDM model cannot fit $z > 0.5$ high-redshift SNe Ia data very well. Despite very small discrepancies in the parameter $\Omega_{k}$ under different models, our Universe is very close to be spatially flat. Considering that the results of the parameter $w \neq -1$ under different models in Table~\ref{table1}, future studies of the \textquote{phantom} dark energy ($w < -1$) and the \textquote{quintessence} dark energy ($w > -1$) are necessarily required to explore the perturbations of dark energy beyond the flat $\Lambda$CDM model \citep{2013LRR....16....6A, 2013CQGra..30u4003T}. Also, since two parameters $w_{0}$ and $w_{a}$ are poorly constrained by the MCMC algorithm, future investigations must be conducted to reduce the parameter uncertainties, which can be realized by incorporating the local expansion rate ($H_{0}$) measurements and more recent SNe Ia constraints \citep{2016ApJ...826...56R, 2022ApJ...938..110B}, running more steps for the MCMC algorithm with faster convergence, and exploring the Pantheon$+$ sample \citep{2022ApJ...938..113S} to fit more SNe Ia data.

\begin{acknowledgements}
This research was originally undertaken as part of the 2022-2023 Level 3 Computing Project module at Durham University Department of Physics. The Pan-STARRS1 Surveys (PS1) and the PS1 public science archive have been made possible through contributions by the Institute for Astronomy, the University of Hawaii, the Pan-STARRS Project Office, the Max-Planck Society and its participating institutes, the Max Planck Institute for Astronomy, Heidelberg and the Max Planck Institute for Extraterrestrial Physics, Garching, The Johns Hopkins University, Durham University, the University of Edinburgh, the Queen's University Belfast, the Harvard-Smithsonian Center for Astrophysics, the Las Cumbres Observatory Global Telescope Network Incorporated, the National Central University of Taiwan, the Space Telescope Science Institute, the National Aeronautics and Space Administration under Grant No. NNX08AR22G issued through the Planetary Science Division of the NASA Science Mission Directorate, the National Science Foundation Grant No. AST-1238877, the University of Maryland, Eotvos Lorand University (ELTE), the Los Alamos National Laboratory, and the Gordon and Betty Moore Foundation. In addition, PP would like to express heartfelt gratitude to Prof. Baojiu Li, who offered technical help in building the MCMC algorithm and explained many physical interpretations in cosmology.
\end{acknowledgements}


\bibliography{main}
\bibliographystyle{raa}

\appendix                  

\section{The magnitude-redshift relation and the MCMC corner plot for the $\MakeLowercase{ow}$CDM model}
\label{appendixa}

\begin{figure*}
  \begin{minipage}{0.5\textwidth}
     \centering
     \includegraphics[width=1.0\linewidth]{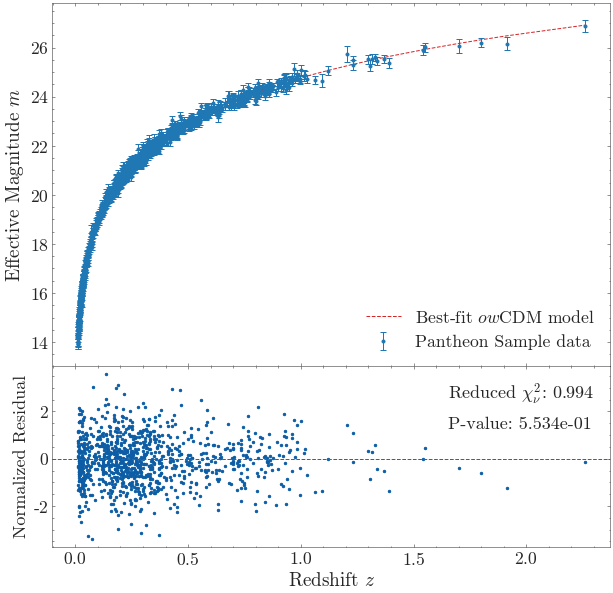}
   \end{minipage}\hfill
   \begin{minipage}{0.5\textwidth}
     \centering
     \includegraphics[width=1.0\linewidth]{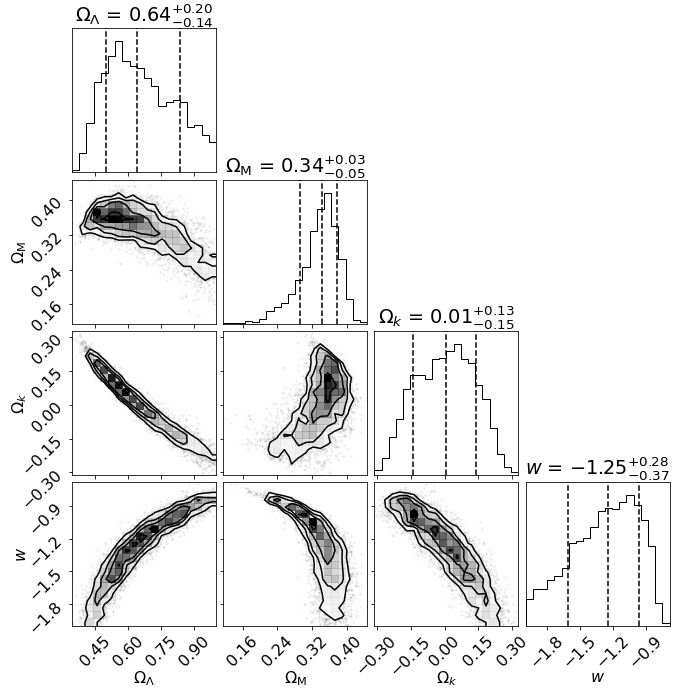}
   \end{minipage}
  \caption{The left panel is the magnitude-redshift relation of the Pantheon Sample fitted by the $ow$CDM model when $\Omega_\mathrm{\Lambda} = 0.639^{+0.196}_{-0.140}$ and $\Omega_\mathrm{M} = 0.343^{+0.035}_{-0.051}$. The normalized residual plot is included underneath to help visualize the fitting ability of the $ow$CDM model. The right panel is the MCMC corner plot for the $ow$CDM model. The best-fit cosmological parameters ($\Omega_\mathrm{\Lambda}, \Omega_\mathrm{M}, \Omega_{k}, w$) and their uncertainties are well constrained by running the MCMC algorithm for 2,000 steps with valid contour plots.}
  \label{figure_a1}
\end{figure*}

\section{The MCMC corner plots for the $\MakeLowercase{o}$CDM, flat $\MakeLowercase{w}$CDM, and flat $\MakeLowercase{w_{0}w_{a}}$CDM models}
\label{appendixb}

\begin{figure*}
   \begin{minipage}{0.5\textwidth}
     \centering
     \includegraphics[width=1.0\linewidth]{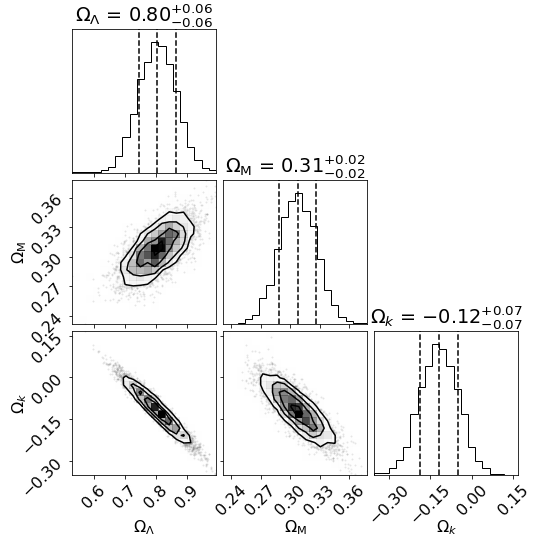}
   \end{minipage}\hfill
   \begin{minipage}{0.5\textwidth}
     \centering
     \includegraphics[width=1.0\linewidth]{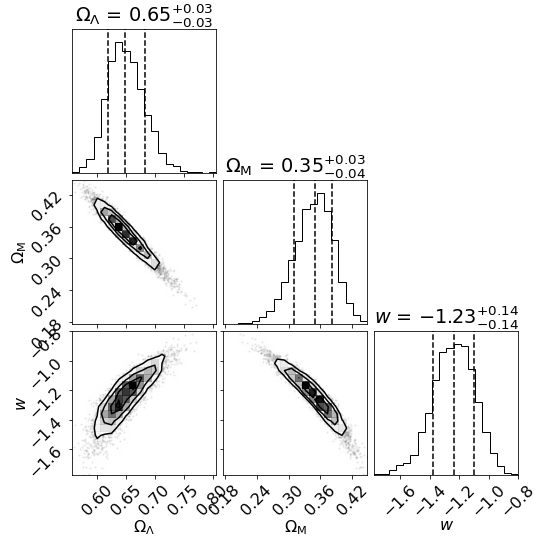}
   \end{minipage}
   \caption{The MCMC corner plots for two models, where the left panel is for the $o$CDM model and the right panel is for the flat $w$CDM model. The best-fit cosmological parameters ($\Omega_\mathrm{\Lambda}, \Omega_\mathrm{M}, \Omega_{k}, w$) and their uncertainties in two corner plots are determined by running the MCMC algorithm for 2,000 steps. All parameters are able to construct valid and well-constrained contour plots, although the cosmological constant $\Omega_\mathrm{\Lambda}$ in the left panel is not consistent with the literature results.}
   \label{figure_b1}
\end{figure*}

\begin{figure*}
  \centering
  \includegraphics[width=1.0\linewidth]{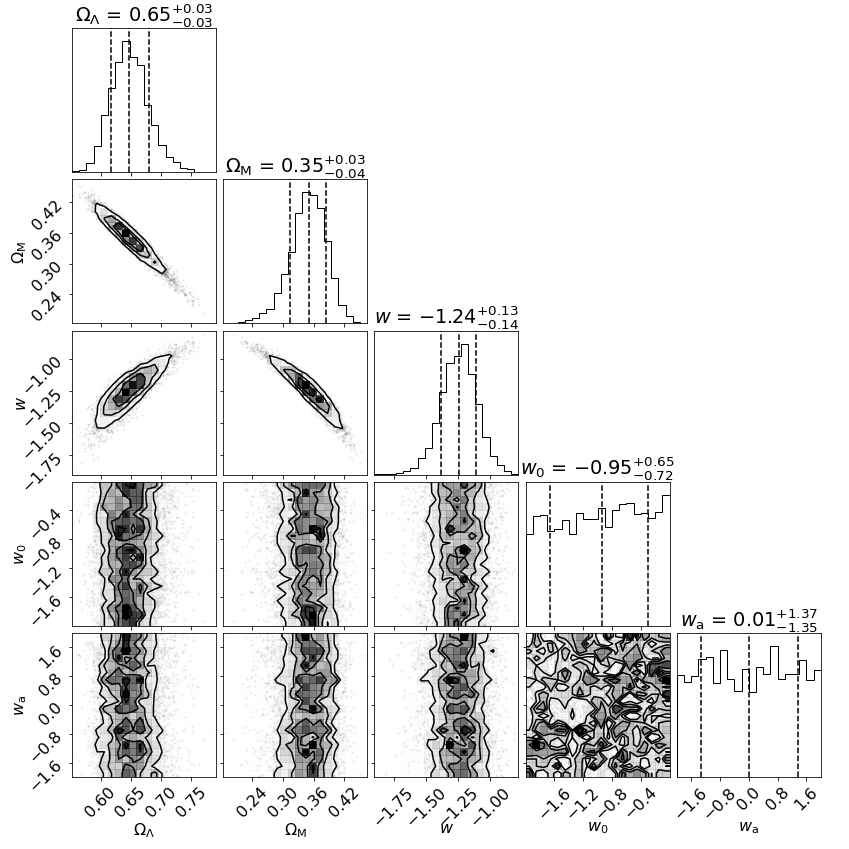}
  \caption{The MCMC corner plot for the flat $w_{0}w_{a}$CDM model. The best-fit cosmological parameters ($\Omega_\mathrm{\Lambda}, \Omega_\mathrm{M}, w, w_{0}, w_{a}$) and their uncertainties are determined by running the MCMC algorithm for 2,000 steps. It is obviously seen that two parameters $w_{0}$ and $w_{a}$ are still poorly constrained in this corner plot, which indicates again that future improvements to the MCMC algorithm are necessarily required.}
  \label{figure_b2}
\end{figure*}

\label{lastpage}

\end{CJK*}
\end{document}